\documentclass[twocolumn]{aastex631}
\usepackage{xspace}
\usepackage{mdframed}
\usepackage{enumerate}
\usepackage{soul}
\usepackage{mathtools}
\newcommand{\tninty}{{$T_{\rm 90}$}\xspace}
\newcommand{\swift}{{\it Swift}\xspace}
\newcommand{\fermi}{{\it Fermi}\xspace}
\newcommand{\gbm}{{\it Fermi}/GBM\xspace}
\newcommand{\bat}{{\it Swift}/BAT\xspace}

\newcommand{\keV}{{\rm keV}}

\newcommand{\s}{{\rm ~s}}

\newcommand{\Ep}{$E_{\rm p}$\xspace}

\newcommand{\sw}[1]{\texttt{#1}}
\newcommand\T{\rule{0pt}{4ex}}

\begin{document}

\title[FermiML]{
Diversity in \textit{Fermi/GBM} Gamma Ray Bursts:  New insights from Machine Learning}
\author[0000-0001-9868-9042]{Dimple}\email{dimplepanchal96@gmail.com}
\affiliation{Chennai Mathematical Institute, Siruseri, 603103 Tamilnadu, India.}
\affiliation{Aryabhatta Research Institute of Observational Sciences, Manora Peak, Nainital-263002, India.}
\author[0000-0003-1637-267X]{K. Misra}
\affiliation{Aryabhatta Research Institute of Observational Sciences, Manora Peak, Nainital-263002, India.}
\author[0000-0002-6960-8538]{K. G. Arun}
\affiliation{Chennai Mathematical Institute, Siruseri, 603103 Tamilnadu, India.}

\begin{abstract}
Classification of gamma-ray bursts (GRBs) has been a long-standing puzzle in high-energy astrophysics. Recent observations challenge the traditional short vs. long viewpoint, where long GRBs are thought to originate from the collapse of massive stars and short GRBs from compact binary mergers. Machine learning (ML) algorithms have been instrumental in addressing this problem, revealing five distinct GRB groups within the \bat light curve data, two of which are associated with kilonovae (KNe). In this work, we extend our analysis to the \gbm catalog and identify five clusters using unsupervised ML techniques, consistent with the \bat results. These five clusters are well separated in fluence-duration plane, hinting at a potential link between fluence, duration and complexities (or structures) in the light curves of GRBs. Further, we confirm two distinct classes of KN-associated GRBs. The presence of GRB~170817A in one of the two KNe-associated clusters lends evidence to the hypothesis that this class of GRBs could potentially be produced by binary neutron star (BNS) mergers. The second KN-associated GRB cluster could potentially originate from NS-BH mergers. Future multimessenger observations of compact binaries in gravitational waves (GWs) and electromagnetic waves can be paramount in understanding these clusters better.
\end{abstract}


% \linenumbers
\section{Introduction} 
\label{sec:intro}
Over five decades after their discovery, the origins of gamma-ray bursts (GRBs) remain a widely debated topic within the astronomy community. The bimodal distribution of GRB duration within the {\it BATSE} data hinted at two classes of GRBs \citep{Kouveliotou_1993}. Based on the duration, \tninty (the duration over which 90\% of the energy in prompt emission is released), GRBs are broadly divided into two classes: short and long GRBs, with a boundary at \tninty of 2 sec \citep{Fishman_1995}. Over the last decades, a broad consensus has emerged that long-duration GRBs are associated with the core-collapse of massive stars \citep{Bloom_2002,Hjorth2003,Woosley2006,Cano2017}, while short-duration GRBs likely arise from a distinct phenomenon, attributed to the mergers of compact objects \citep{Paczynski1986, Berger2013, Tanvir2013, Abbott_2017, Troja2019}.

Despite the initial success of the classification based on \tninty, with data from various GRB detectors covering different energy bands, the limitations of this classification are widely acknowledged~\citep{Fynbo_2006,zhang2009,Qin2013}. An overlap between two populations is also observed, where it is hard to determine if a burst with mixed characteristics belongs to the short or long class \citep{Fynbo_2006,Dimple_2022,Troja2022, Becerra_2023}. Recent observations suggest further diversity within these classes. For example, some short GRBs exhibit properties that deviate from the merger model and vice versa \citep{Bromberg2013,Dimple2022b,Dimple_2023b, Petrosian_2024}. Long GRBs~211211A \citep{Rastinejad2022, Troja2022, Gompertz2022, Yang2022} and 230307A \citep{Dichiara_2023,Sun_2023,Levan_2024} have observational evidence of a kilonova (KN) association in their optical light curves. Similarly, the short GRB~200826A was found to be associated with a supernova (SN) hinting at its origin from a collapsar \citep{Ahumada2021, Rastinejad2022}. These cases hint at a more complex picture, suggesting that the progenitors responsible for GRBs might not be as straightforward as previously believed.

Recent advances in machine learning (ML) and artificial intelligence (AI) are fostering a more detailed understanding of GRBs \citep{Chattopadhyay_2017, Acuner_2018, Christian_2020, Salmon_2022, Steinhardt2023, Keneth_2023, Mehta_2024, negro2024prompt, Zhang_2024, Kumar_2024, Zhu_2024}. When applied to the analysis of prompt emission light curves, these techniques have revealed hidden patterns within the GRB population, suggesting the existence of subclasses beyond the traditional short--long GRB classification \citep{Christian2020, Steinhardt2023, Keneth_2023,Dimple_2023}.

An ML-based analysis of the prompt emission light curves using \bat data has revealed five distinct clusters within the GRB population, which might represent distinct progenitors~\citep{Dimple_2023}. Among these clusters, the GRBs associated with KNe were found to lie in two distinct clusters, suggesting that they may have been produced by different progenitors. The clusters could correspond to binary neutron star (BNS) and neutron star–black hole (NS-BH) mergers as GRB progenitors or subclasses of either of them. These findings, if confirmed, can have profound implications for our understanding of GRBs and compact binaries. One way to validate these findings is to wait for a sizeable number of multimessenger GRBs such as GW170817. Given the current sensitivities of gravitational wave (GW) detectors, this may take a few years. Another way forward is to validate these findings using data from other datasets, which constitutes the theme of this paper.

One such dataset is the \fermi Gamma-ray Burst Monitor (GBM) catalog \citep{von_2020}, which compiles data of GRBs detected by the \gbm. \gbm data covers a wide energy spectrum, ranging from a few keVs to several MeVs, and includes nearly twice as many GRBs compared to the \bat catalog \citep{Lien_2016}. Further, \gbm observed the multimessenger event GW170817 \citep{Goldstein2017}, originating from a BNS merger \citep{Abbott2017}. Studying the GRBs in this catalog can provide a more comprehensive understanding of GRB progenitors.

This paper presents a comprehensive study of the GRB population using ML techniques on \gbm prompt emission light curves. Previous studies by \citet{Acuner_2018} and \citet{Mehta_2024} employed ML techniques on the \gbm data to classify GRBs using spectral parameters. However, our approach differs from these works. We utilize light curves across different energy bands and apply unsupervised ML algorithms to cluster GRBs based on similarities and dissimilarities in their light curves. The paper is structured as follows: Section \ref{methods} describes the methodology employed for the data analysis and clustering of GRBs. Section \ref{results} illustrates the outcomes of our ML analysis and discusses their implications. Lastly, Section \ref{summary} summarizes the study, highlighting the main findings and their prospective influence on future research.

\section{Methodology }
\label{methods}
We compiled the prompt emission data files for the GRBs observed from July 2008 to August 2023 from the \gbm catalog, inclusive of the time-tagged event (TTE) and response files for the detector configuration sourced from the \gbm archive\footnote{\url{https://heasarc.gsfc.nasa.gov/W3Browse/fermi/fermigbrst.html}}. This section outlines the methodology used to analyze the GRB data and create the clustering through ML algorithms.

\begin{figure*}
    \centering
    \includegraphics[height=6.6cm,width=\columnwidth]{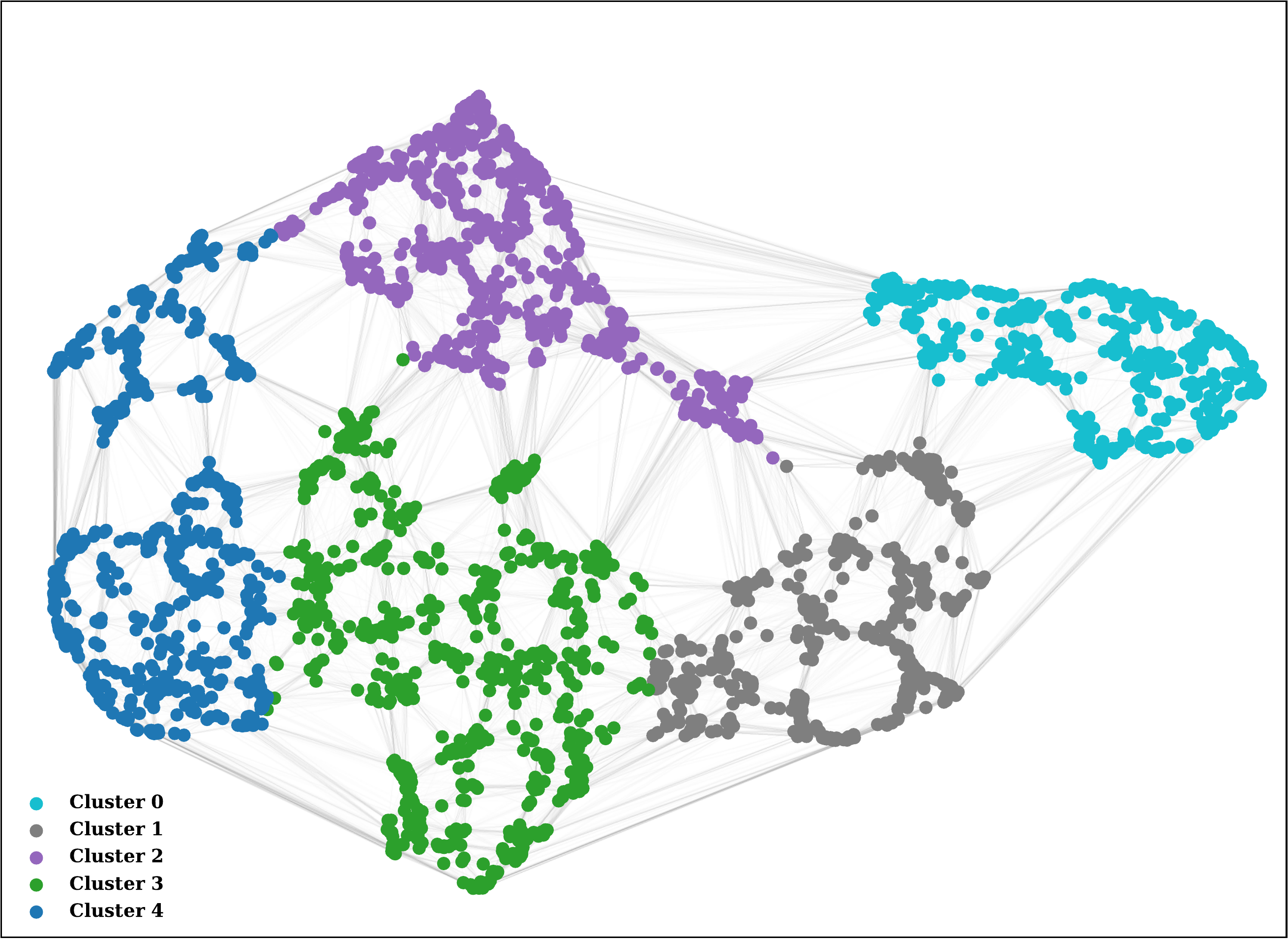}
    \includegraphics[height=6.6cm,width=\columnwidth]{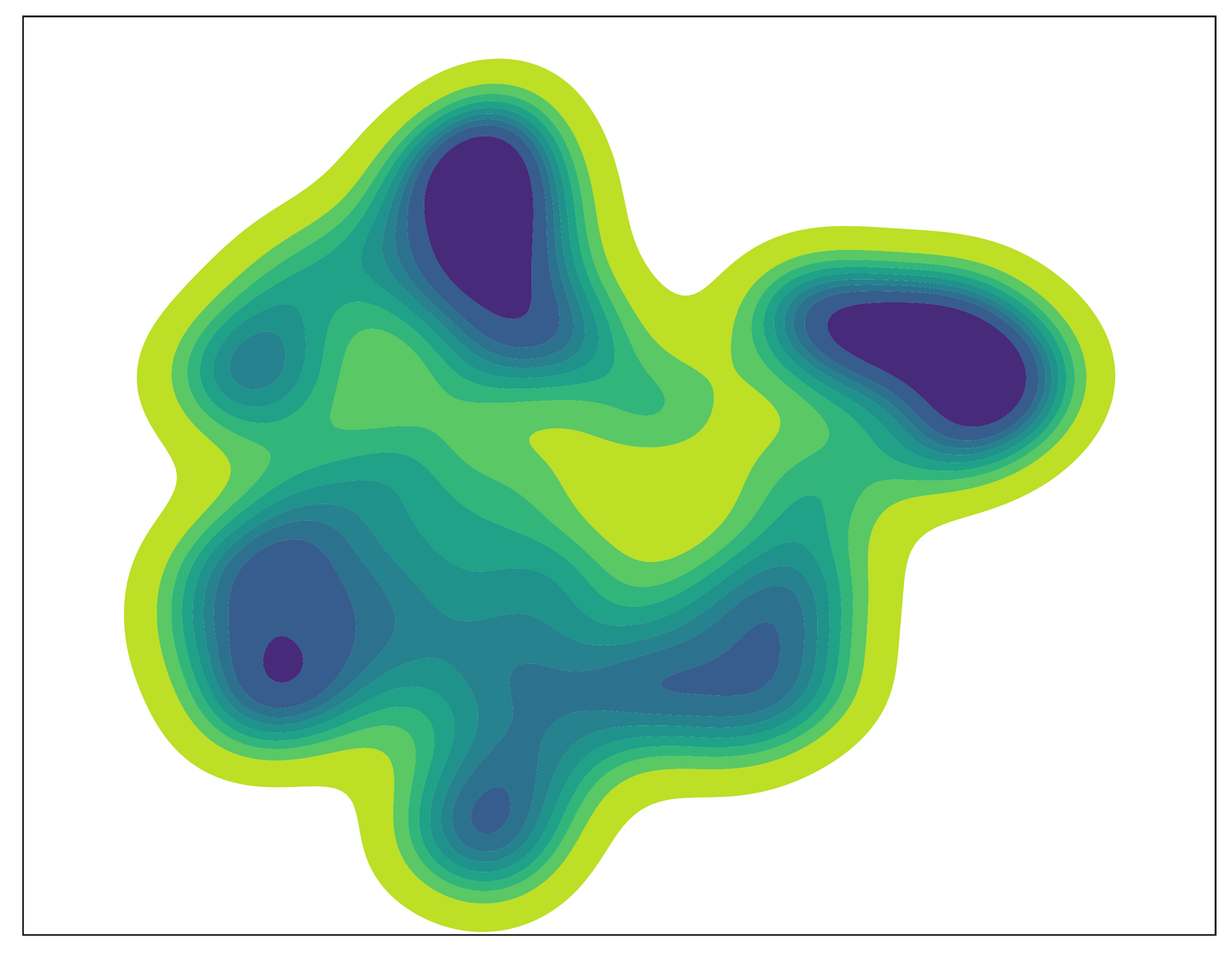}
    
    \caption{This figure visualizes the embeddings obtained using PCA-UMAP on the prompt emission light curves (in three energy bins) of 3516 GRBs. \textbf{Left:} Connectivity map where each point represents a GRB and edges represent how it is connected to other bursts. The bursts are color-coded with the clusters identified by \sw{AutoGMM}. \textbf{Right:} The respective density map, the deeper the shade of blue, the higher the density of GRBs, meaning more GRBs are found in that region. The density map is consistent with the groups identified by \sw{AutoGMM}.}
    \label{fig:connectivity_maps}
\end{figure*}

\subsection{Data Reduction Pipeline}
We designed a systematic pipeline implemented in \sw{Python} to extract light curves in various energy bands and prepare the data for the ML algorithms. The pipeline uses the \sw{GbmDataTools} package \citep{GbmDataTools} to access and analyze GBM data. This pipeline automates several key steps, as listed below:

\begin{enumerate}[a)]
    \item  Detector selection: The process begins by estimating the angles between each detector and the direction of the burst. It then identifies the Sodium Iodide (NaI) detector with the smallest angle to the burst and selects the corresponding Bismuth Germanium Oxide (BGO) detector. The data from these detectors are used for further analysis.
    
    \item Light curve extraction: Light curves are extracted at different energy bands: 8--50 keV and 50--300 keV for the NaI detector and 300--1000 keV for the BGO detector. The pipeline uses a temporal binning of 16 milliseconds from TTE data files using the \sw{TTE} function within \sw{GbmDataTools}. 
    
    \item Background subtraction: The background signal is fitted and subtracted from the corresponding light curves. The \sw{BackgroundFitter} function in \sw{GbmDataTools} is used for background fitting. The resulting light curves are used for subsequent processing. 
    
    \item Data denoising: Before using the light curves for ML analysis, unwanted noise, which could obscure the true variations in the data, is removed. Wavelet analysis is employed for denoising. The optimal wavelet and decomposition level are chosen to minimize the mean squared error between the original and denoised data. This process effectively reduces noise in the time series data. 
    
    \item Dataset preparation for ML algorithms: The background-subtracted light curves are normalized with the fluence to mitigate the potential algorithmic distractions caused by the typical disparity in fluence between long and short GRBs. Following normalization, a standardization procedure is applied to ensure uniformity in the length of the light curves. All light curves are shifted to a common starting point and padded with zeros. Subsequently, the normalized and standardized light curves are concatenated, forming a cohesive time-series dataset with a shared axis and consistent lengths across different GRBs. Finally, a discrete-time Fourier transform (DTFT) is employed on the concatenated data set to preserve the time delay information between various light curves. 
\end{enumerate}

After standardization, we could recover 3349 GRBs from the \gbm catalog. For the ML algorithm, we used the data in three energy bands, two energy bands of NaI (8--50 \keV and 50--300 \keV) and one energy band of BGO (300--1000 \keV) as NaI is more sensitive in the lower energy band, and BGO is more sensitive in the high energy bands. We stored the standardized data as a matrix sized [3349 x 81798], where each row represents a GRB and includes its light curves across three energy channels binned at 16 ms intervals.

\subsection{Clustering using unsupervised learning}
We employed Uniform Manifold Approximation and Projection (UMAP; \citealt{McInnes2018}) with Principal Component Analysis (PCA) for dimensionality reduction. PCA-UMAP (discussed in detail in \citealt{Dimple_2023}) allows the analysis of high-dimensional data in a lower-dimensional space while preserving its underlying structure. For that, we first extracted the first 500 principal components using PCA, capturing 99\% of the data variance. Then UMAP is used to reduce the dimensionality of the data further using topology. UMAP builds a network of connections between data points based on similarity and dissimilarity. This network reflects the inherent structure of the data, capturing local and global relationships. Then, it projects the data to lower-dimensional space while maintaining these relationships. The algorithm relies on two key parameters: \sw{n\_neighbors}, which limits the size of the local neighborhood considered to build the topological structure, and \sw{min\_dist}, which controls how tightly packed nearby points are. By adjusting these parameters, we obtained a more precise and insightful representation of our data in lower dimensions. Specifically, we set the value for \sw{n\_neighbors} at 25 and \sw{min\_dist} at 0.01.

We generated the connectivity and density maps to further analyze the resulting low-dimensional representation. Connectivity maps visualize the probability of a connection between two data points, essentially depicting the aforementioned network in a two-dimensional format. On the other hand, density maps, along with the 2-dimensional embeddings, offer additional insights into the data structure. These maps help to understand the distribution and concentration of data points in a low-dimensional space. Finally, we identified clusters within the two-dimensional embedding using the \sw{AutoGMM} module \citep{Athey2019}, which applies a Gaussian Mixture Model (GMM) to automatically determine the optimal number of clusters present in the data. \sw{AutoGMM} utilizes parameters such as the minimum and maximum number of clusters for cluster identification. For our analysis, we set the minimum number of clusters to 2 and the maximum to 7. This range was chosen based on density maps, which visually guide the expected number of clusters in embeddings and help prevent overfitting. We used the Bayesian Information Criterion (BIC) to determine the best classification, maintaining default settings for other parameters.

\begin{figure}
    \centering
    \includegraphics[width=\columnwidth]{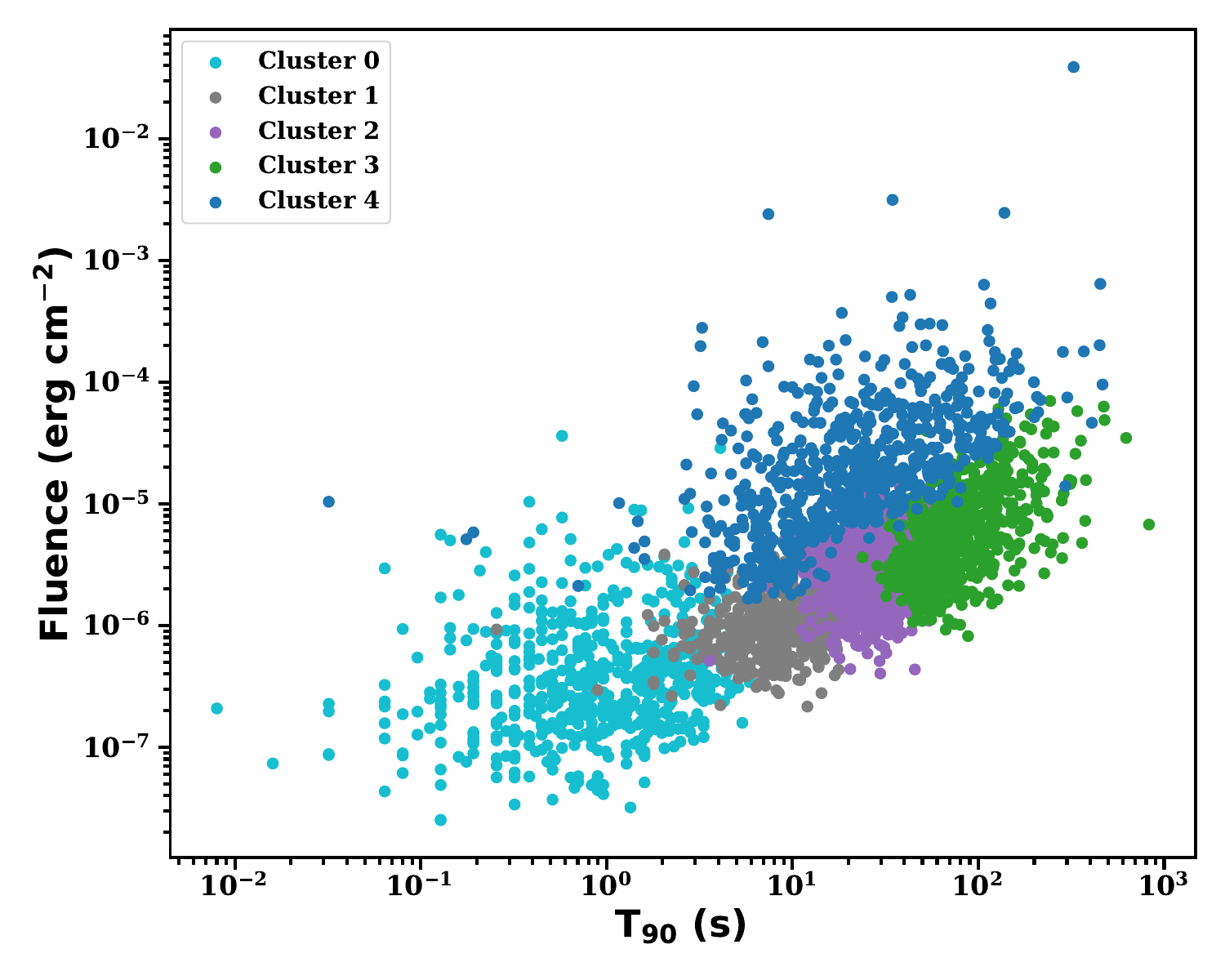}
    \caption{GRB locations on fluence--\tninty plane: GRBs are color coded based on distinct clusters identified by \sw{AutoGMM}. Notably, these clusters exhibit clear separation in the fluence-duration plane, suggesting that light curve features have dependence on GRB fluence and duration.}
    \label{fig:fl_vs_t90}
\end{figure}

\begin{figure*}
\centering
\includegraphics[width=0.8\linewidth]{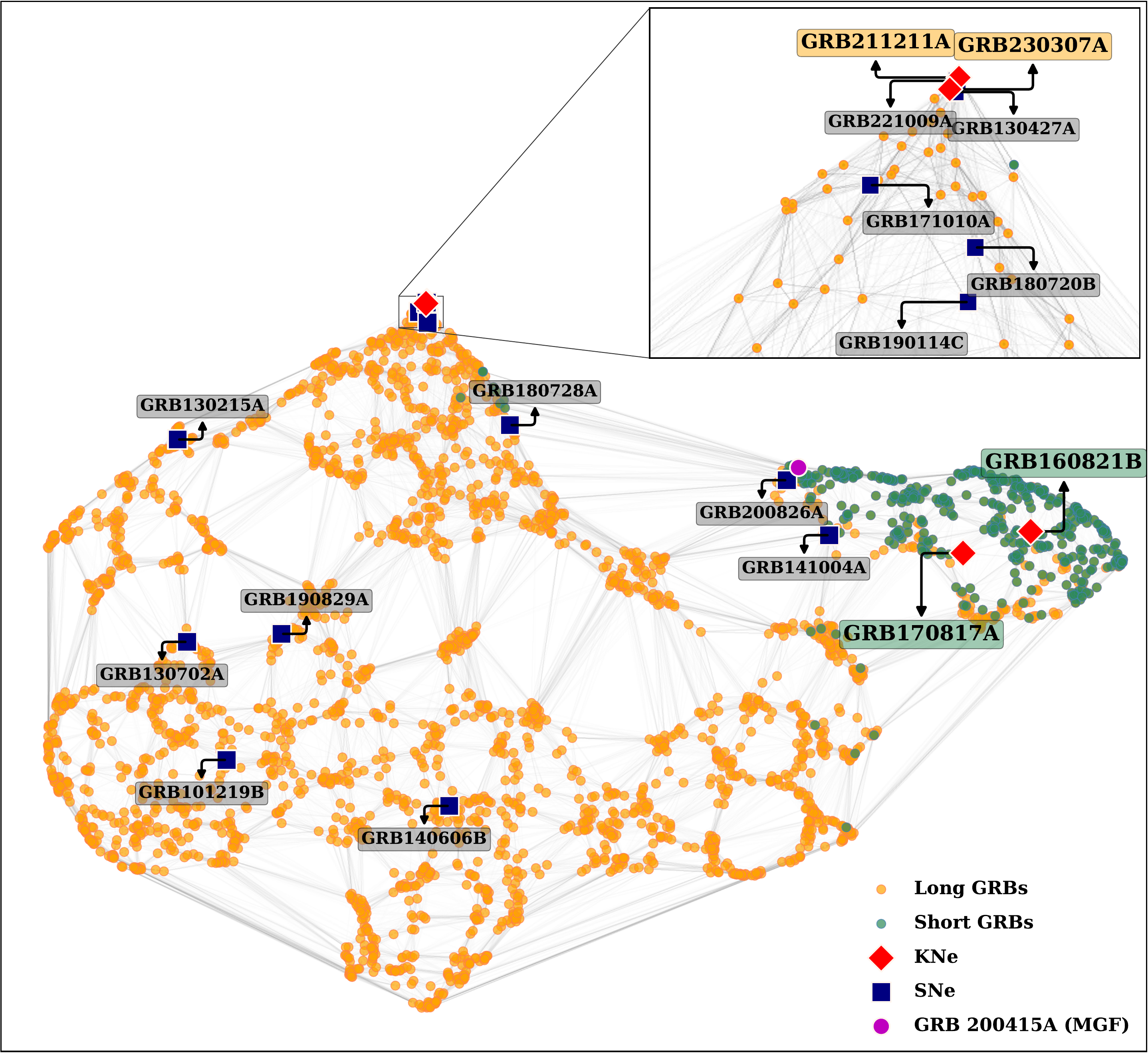}
    \caption{The connectivity map color coded as traditional long and short duration GRBs based on their \tninty duration. KN and SN-associated GRBs are shown with red diamonds and blue squares, respectively, on the embeddings. KN-associated GRBs are labelled with their names and respective colors as per their traditional classification. Short and long-duration KN-associated GRBs are well separated into two distinct groups. SN-associated GRBs are scattered across the map with an unclear clustering pattern. Notably, several SN and KN-associated GRBs lie at the top of cluster 2. The inset offers a closer view, revealing that GRBs~211211A and 230307A are adjacent, suggesting similarity in their light curves. Additionally, some bright SNe are also situated nearby labeled with their names in grey color.}
    \label{kn_sn_candidates}
\end{figure*}

\section{Results and discussion}
\label{results}
We obtained the two-dimensional embeddings using the method described in Section~\ref{methods}. The left panel of Figure~\ref{fig:connectivity_maps} shows these embeddings with connectivity maps color-coded with the clusters identified by \sw{AutoGMM}. The right panel of the figure displays the respective density maps, providing additional information about the density of GRBs in the map. 

\subsection{Distinct groups within GRBs}
Our method identifies five distinct groups within the embedding, evident from the density plots (see Figure~\ref{fig:connectivity_maps}). This is remarkably similar to the findings of \citet{Dimple_2023}, where five distinct clusters were identified using the same method on the \bat data. It is also interesting to note that, based on spectral parameters that govern the emission mechanism, \citet{Acuner_2018} showed the existence of five clusters in the \gbm data. However, we do not find an obvious overlap of the clusters we identify here with those of \citet{Acuner_2018}. Besides the \bat and \gbm datasets, the {\it BATSE} data also revealed five clusters in a study by \citet{Chattopadhyay_2017}. Given the differences in the detectors and their energy bands, among other things, the match in the number of clusters is very intriguing.  

The consistency of five distinct clusters across multiple datasets and methodologies suggests that the GRB population exhibits diversity beyond just two classes. To further investigate these clusters, we examined their prompt emission properties. We observed that these clusters exhibit minimal differences in their spectral properties, such as \Ep and $\alpha$ (See Figure~\ref{prompt_emsn_prop} in Appendix~\ref{properties}). However, a clear trend in duration and fluence can be observed for these clusters. The separation of clusters based on their duration is expected, but fluence also contributes to distinguishing the clusters. For instance, clusters 2 and 4 overlap in their duration values but differ in fluence values, suggesting that both fluence and duration have influenced the clustering. To further explore these clusters, we plotted the GRBs from the five clusters on the fluence--\tninty plane (Figure~\ref{fig:fl_vs_t90}), and interestingly, they are well separated in this plane. This allows the identification of various classes: short-duration bursts with low fluence (cluster 0), intermediate-duration bursts with low fluence (clusters 1 and 2), long-duration bursts with low fluence (cluster 3), and intermediate to long-duration bursts with high fluence (cluster 4).

It is important to note that even though we removed the fluence information during data standardization, fluence still plays a role in clustering. This suggests that fluence information is conveyed through the light curves themselves, either through duration or structures within the light curves. This implies that the complexities of light curves may depend on brightness, which is also suggested by \citet{Stern_1997}. Though the correlation between fluence and duration is well known \citep{Horvath_2001}, our results hint that there can be a correlation between light curve shapes and brightness, which may have a deeper connection with the physics of the central engines of GRBs. A detailed study of the implications of this correlation for the central engines of GRBs may be the subject of future work. 

Given that our clustering is based solely on the prompt emission light curves, these distinct classes may represent different progenitors, central engines, emission mechanisms, or a combination of these. To date, evidence from optical bumps (SNe/KNe) suggests that the collapse of massive stars (collapsars) or the merging of compact objects (NS/BH/WD) are two possible scenarios for the origin of GRBs. We locate all the candidates for KNe and SNe on the connectivity maps. SN-associated GRBs are spread across the map, as evident from Figure~\ref{kn_sn_candidates}. Interestingly, some of these candidates are adjacent to KN-associated GRBs. SN-associated GRBs~221009A, 130427A, 
lie very close to KN-associated GRBs~211211A and 230307A. Further, GRB~200826A, a short GRB associated with an SN, lies in the cluster shared by KN-associated GRBs, GRBs~170817A and 160821B along with one SN-associated GRB~141006A.
\citet{Li_2023} also reported similarity in the temporal and spectral properties of some of the KN- and SN-associated GRBs. This indicates that despite differences in the progenitors, they might have similar central engines and/or emission mechanisms, resulting in similar light curves. It is possible that collapsars and compact binary mergers may result in the formation of a central engine, such as a hyperaccreting black hole or a magnetar, which could be responsible for launching a relativistic jet \citep{Duncan_1992, Usov_1992, Popham_1999, Narayan_2001, Metzger2019}.

We also located the magnetar giant flares (MGFs), which mimic the short GRBs. We have only one MGF in our sample, 200415A (see Figure~\ref{kn_sn_candidates}), which is very close to GRB~200826A. 

\subsection{Two distinct populations of KN-associated GRBs}
Next, we look for the locations of the KNe-associated GRBs in the \gbm map. This is motivated by the findings of \citet{Dimple_2023}, who argued for two distinct populations of GRBs associated with KNe in the \bat data. Our KN sample comprises four confirmed cases of KN-associated GRBs: 160821B \citep{Kasliwal2017,Lamb2019,Troja2019}, 170817A \citep{Abbott_2017, Goldstein2017,Valenti_2017}, 211211A \citep{Rastinejad2022,Troja2022}, and 230307A \citep{Dichiara_2023,Sun_2023,Levan_2024}.

The results are presented in Figure~\ref{kn_sn_candidates}, which interestingly confirms the existence of two clusters of KNe-associated GRBs based on \bat. Unlike \bat, \gbm has observed GRB~170817A, known to be a BNS merger from GW observations \citep{Abbott_2017} with a confirmed detection of a KN. Therefore, we find evidence favouring the conjecture of \citet{Dimple_2023} that the low fluence, small duration cluster may correspond to BNS mergers. Similarly, GRBs~170817A and 160821B are closely located in the same cluster, implying they could possibly originate from the same kind of progenitor systems. \citet{Dimple_2023} argued that the second KNe-associated cluster could potentially be arising from NS-BH mergers. 
As GRB~230307A is not detected by \bat, the proximity of GRBs~211211A and 230307A in the \gbm map suggests the similarity in their light curves and the progenitors as also suggested by \citet{Peng_2024}. Both bursts exhibit three consistent emission episodes: a precursor, a main burst, and an extended emission phase (see figure 1 in \citealt{Peng_2024}). The spectral lags for corresponding episodes in both GRBs are nearly zero, and the minimum variability time scales are comparable \citep{Peng_2024}. They are also lying in a close vicinity in the low-dimensional embeddings obtained using deep learning on \gbm data \citep{negro2024prompt}. However, multimessenger observation of an NS-BH system is necessary to confirm the prediction that GRBs in this cluster are products of NS-BH mergers. For the typical NS masses, the BH must be of mass $\sim 3M_{\odot}$ for the NS-BH to produce a GRB. Recent observation of an NS-BH merger GW230529~\citep{GW230529} with a BH between $2.5-4.5M_{\odot}$ indicates there may be a population of this low mass BH in the universe. With all these, the first multimessenger NS-BH merger will revolutionize our understanding of compact objects and GRBs. 

However, the presence of SNe-associated GRBs in this cluster complicates this paradigm. As SNe-associated GRBs are likely to be associated with massive stellar explosions, why their light curves share features similar to those of the KNe-associated GRBs is unclear. If, for instance, the mass of the central engine plays an important role, then SN explosions that leave behind BHs of mass $2.5-4.5M_{\odot}$ may share similar features in the light curves as that of NSBH mergers whose BH mass also lies in the same range. Admittedly, this is only a speculation at present that needs a careful and dedicated study in the future.

\section{summary and future aspects}
\label{summary}
ML algorithms are slated to play an important role in classifying GRBs, which is an open issue. Our current study looked into the classification of \gbm GRBs using the latest catalog. Listed below are the three key takeaways from this work.
\begin{enumerate}
    \item An ML-based classification of the GRB light curves of \gbm suggests five distinct clusters of GRBs. Independent analyses in the case of \bat and {\it BATSE} also find the number of clusters to be five, potentially hinting there is something unique and fundamental about this number. 
    \item Of these five clusters, two distinct clusters contain KNe-associated GRBs. One cluster contains GRB~170817A, which is associated with a BNS merger. This strengthens the case for these two clusters to be associated with BNS and NS-BH mergers as proposed by \citet{Dimple_2023}. However, these clusters are surrounded by some SNe-associated GRBs, which may possibly indicate the universality of the central engines of GRBs regardless of their progenitors.
    \item Despite the use of fluence-normalized light curves, the ML-based clusters are well-separated in the fluence-\tninty plane, hinting that fluence also influenced the clustering. This indicates that the light curve shapes and duration may have dependencies on the brightness (also suggested by \citet{Horvath_2001} and \citet{Stern_1997} earlier), and is worth exploring.
\end{enumerate}

We next discuss the implications of these findings. Even though the \fermi and \swift datasets contain mostly distinct events, with only a small overlap, their clustering patterns are remarkably similar (see Appendix \ref{crossmatching} and Figure~\ref{Comparison}). This consistency highlights the robustness of the ML-based classification. Further, KNe-associated GRBs are tightly clustered at two locations in both datasets. These clustering patterns can be used for early identification of such GRBs. This may be achieved by reanalyzing the existing data with the light curve of a new GRB in question and asking whether it lies in one of the two clusters. Such an algorithm can be developed and applied to every GRB, assigning a probability of a KNe association. This should enable dedicated deep searches for KNe counterparts associated with these GRBs, even without a GW observation. This will require developing a statistical framework (and a related analysis pipeline) that will carry out this exercise in a self-consistent manner. Moreover, these predictions will save the telescope time as it is impossible to perform a targeted search of all the GRBs, given the faint and rapid decay nature of KNe. The recent discovery of KNe associated with long GRBs further complicates the search, as previously thought to be exclusive to short GRBs. However, some of these bursts are surrounded by SN-associated GRBs. It remains interesting to follow up on them, as detecting either a SN or a KN is valuable.

Secondly, assuming the two clusters correspond to BNS and NS-BH mergers, the rate of GRBs in these clusters could be translated into rates of BNS and NS-BH mergers under reasonable assumptions about the detector's sky coverage, beaming angle range, etc. This may be compared against the rates of BNS and NS-BH mergers from GW observations to see if these two estimates are consistent \citep{shasvath2024}. As the next-generation GW detectors such as Voyager \citep{2020CQGra..37p5003A}, Cosmic Explorer \citep{evans2021horizon}, and the Einstein Telescope \citep{2012CQGra..29l4013S,kalogera2021generation} that are capable of detecting BNS and NS-BH mergers even at higher redshifts are operational, this method can give us valuable insights about the relation between GRBs and compact binary mergers.
 
Finally, as a caveat, one should note that we are not using redshift-corrected light curves, as the number of GRBs with redshift estimation is limited. In the future, with a large dataset with known redshifts, we can explore how time dilation affects this clustering. Further, to advance this paradigm, it may be beneficial to employ the ML approach on the afterglows of these GRBs and study what we can learn from combining the multiband data.

\section*{acknowledgments}
We extend our sincere thanks to the referee for the valuable comments, which enhanced the quality and clarity of the manuscript. We thank the \gbm and \bat team for providing extensive data through public archives. Dimple thanks P. Sreekumar, Ranjeev Mishra, Siddharth Pritam, Ben Gompertz for fruitful discussions. K.G.A. and Dimple acknowledge the Swarnajayanti grant DST/SJF/PSA-01/2017-18 of the Department of Science and Technology and SERB, India, and support from Infosys Foundation. KM and K.G.A. acknowledge the support from BRICS grant DST/ICD/BRICS/Call-5/CoNMuTraMO/2023 (G) funded by the Department of Science and Technology (DST), India. Dimple acknowledges MCNS for their warm hospitality during her stay in the campus.

\bibliography{main}{}
\bibliographystyle{aasjournal}

\appendix
\counterwithin{figure}{section}
\section{Characteristics of the clusters}
\label{properties}
\begin{figure*}[!h]
\centering
    \includegraphics[width=0.48\columnwidth]{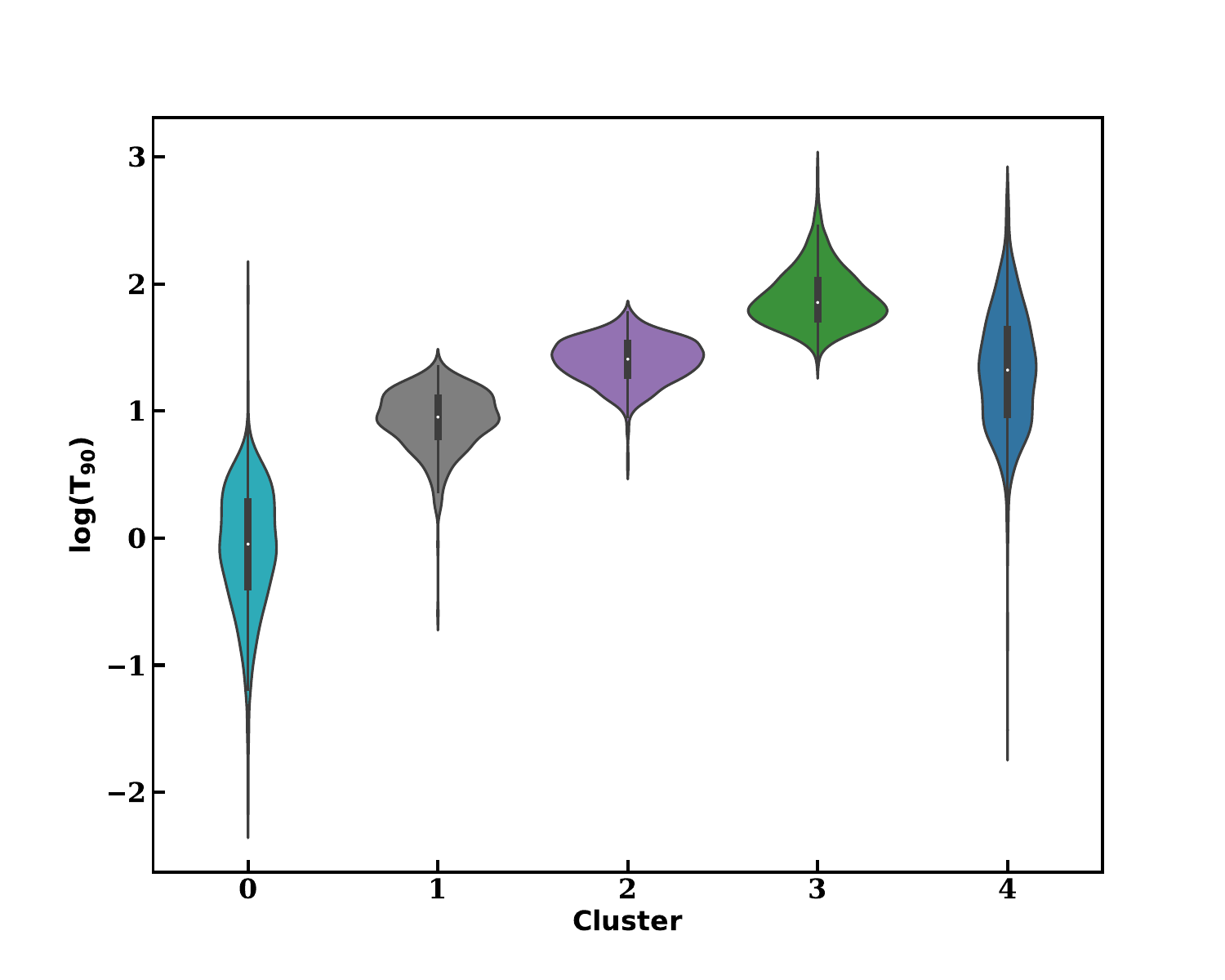}
    \includegraphics[width=0.48\columnwidth]{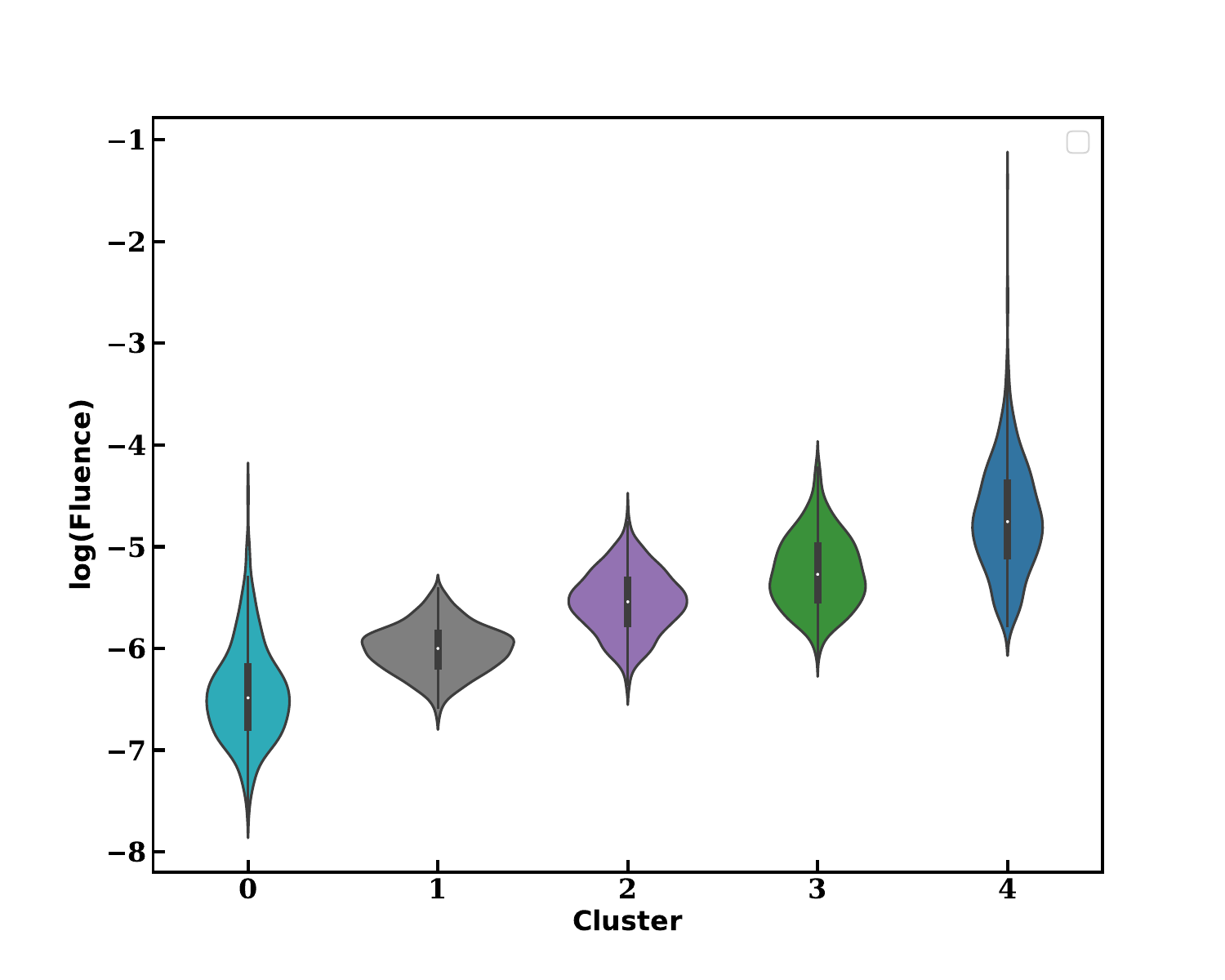}
    \includegraphics[width=0.48\columnwidth]{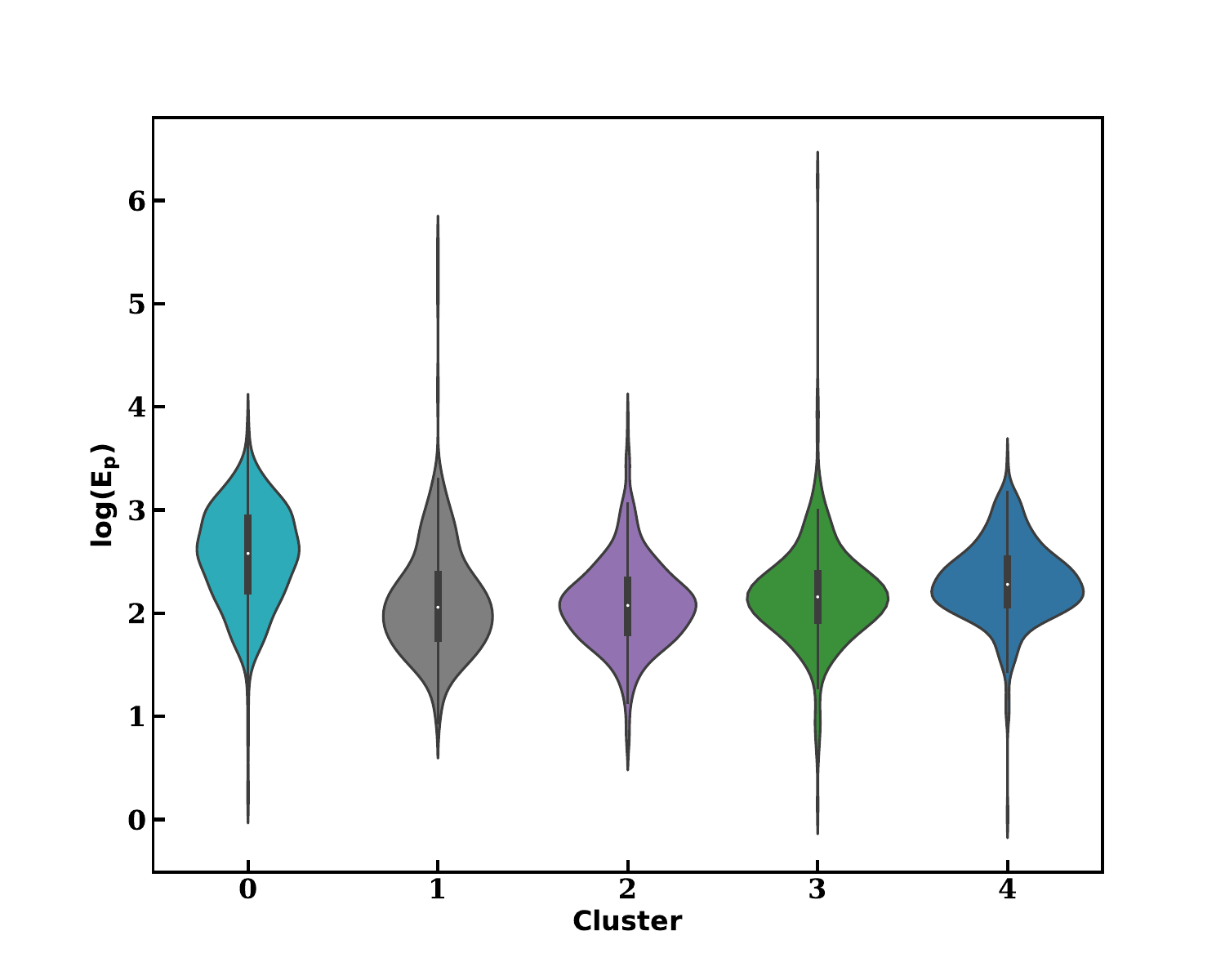}
    \includegraphics[width=0.48\columnwidth]{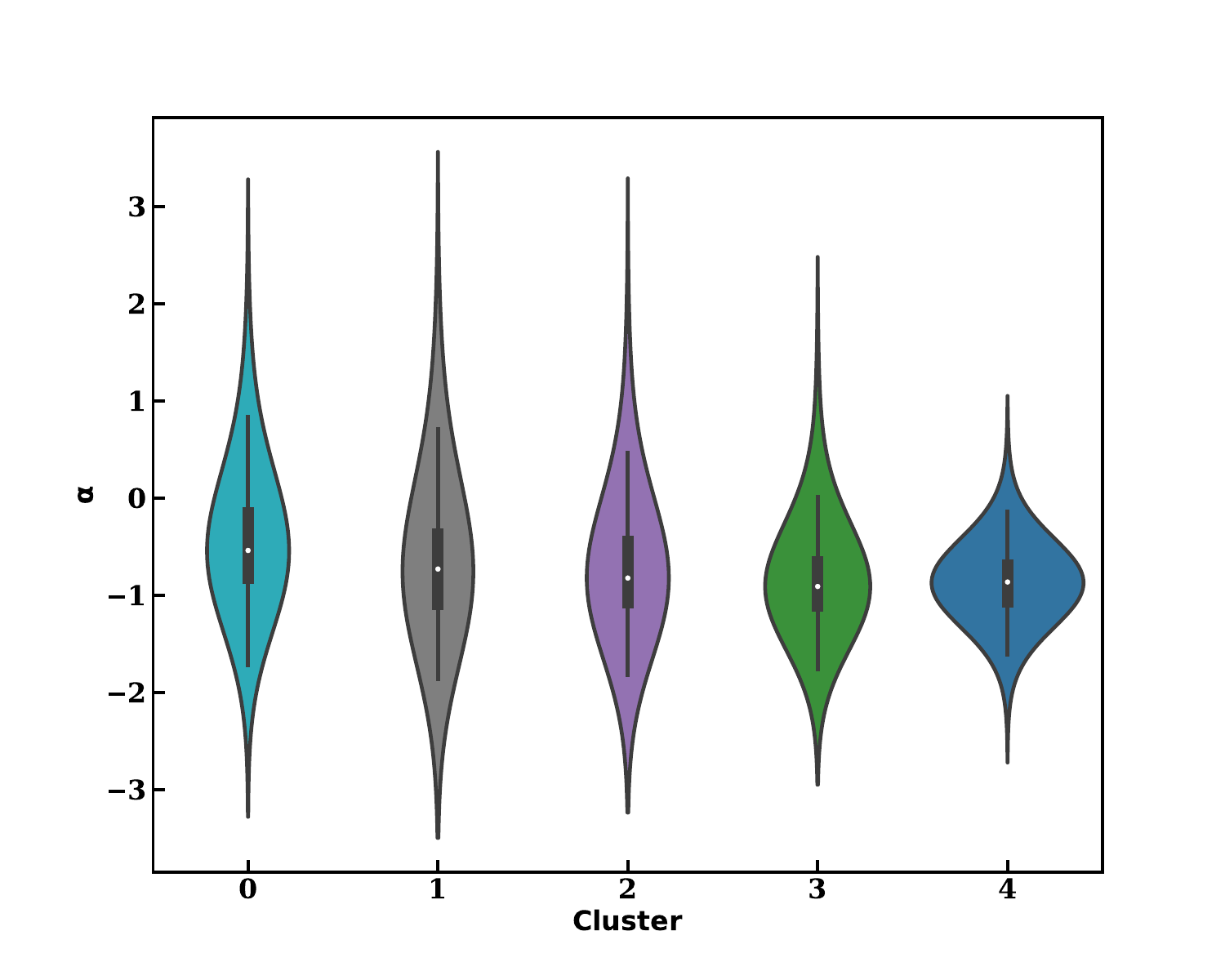}
    \caption{Violin plots for the known GRB properties corresponding to the five groups of GRBs identified by ML algorithms.}
    \label{prompt_emsn_prop}
\end{figure*}

We probe the known properties of GRBs to investigate what these clusters might represent. We took the values of the parameters \tninty, fluence, \Ep and $\alpha$ from the \gbm catalog\footnote{\url{https://heasarc.gsfc.nasa.gov/W3Browse/fermi/fermigbrst.html}}. Figure~\ref{prompt_emsn_prop} shows the violin plots for $T_{90}$ (upper left panel), fluence (upper right panel), and $E_{p}$ (lower left panel), $\alpha$ (lower right panel) for each cluster.

\section{Cross-Matching and Clustering Consistency of GRB Embeddings Across \swift and \fermi Datasets}
\label{crossmatching}
\begin{figure*}[!h]
\centering
    \begin{mdframed}
    \includegraphics[width=\columnwidth,trim={2.8cm 4cm 3.5cm 4cm},clip]{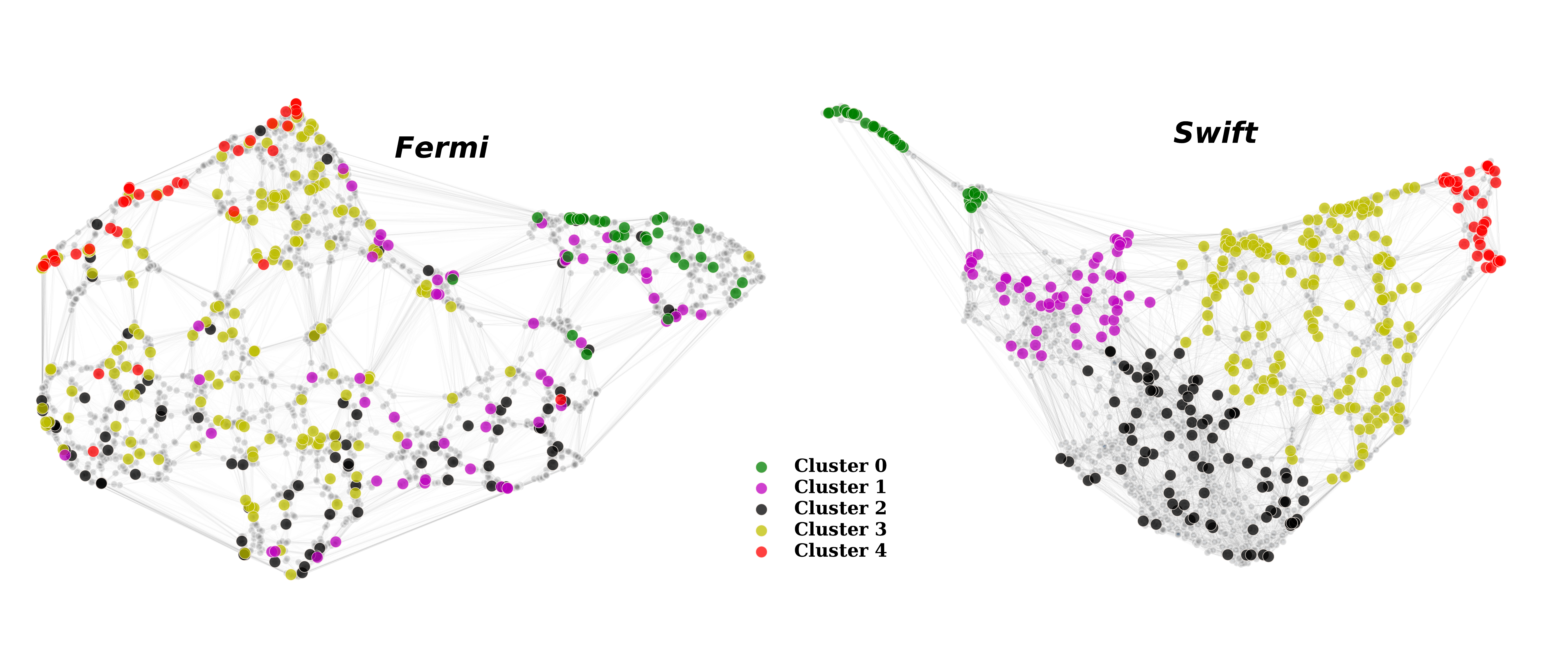}
\end{mdframed}
    \caption{The figure displays the mapping of common GRBs, color-coded by their respective clusters, on the two-dimensional embeddings derived from the \swift and \fermi datasets.   }
    \label{Comparison}
\end{figure*}

We compared the embeddings obtained using the \gbm dataset with those obtained using the \bat dataset. To do this, we first updated our \bat sample (defined in \citet{Dimple_2023} with the GRB light curves observed up to September 2023. We used PCA-UMAP and \sw{AutoGMM} to cluster the bursts in the \bat dataset. For this, we set the value for \sw{n\_neighbors} at 15 and \sw{min\_dist} at 0.01. We identified the clusters using \sw{AutoGMM}, which identified 5 clusters in the dataset as shown in the right panel of Figure~\ref{Comparison}. 
We cross-matched the clusters for common GRBs in the \swift and \fermi catalogs. We compared the cluster numbers (as identified in Figure~\ref{Comparison}) of 375 common bursts and found the cluster number matched for 139 bursts.
It is noteworthy that even those bursts which did not share the same cluster number in both the maps were seen to lie in the neighbouring clusters in most cases we visually inspected. For instance, \swift bursts in Cluster 0 (green) remained in the same cluster in the \fermi map, as seen in the figure. A similar trend can be seen in the case of Cluster 4 (red), which corresponds to the same cluster in the Fermi map, except for a few bursts, which are seen to spread slightly into the nearby clusters. The extent of mixing for other clusters is greater, where some of the bursts in the Fermi map are seen to move to nearby clusters w.r.t. the Swift map.

Potential reasons for these discrepancies may include several factors. One key factor could be the different energy ranges of the light curves used for clustering for two datasets. The two satellites overlap in the lower energy ranges, but \swift does not include the energy bands from 350-1000 keV, which are covered by \fermi. These higher energy bands may contain critical information, ultimately causing differences in the embeddings. Additionally, we used 16ms binning for \gbm light curves, which include more detailed structures. In contrast, the \bat light curves were in 64ms (as we could not recover many light curves with 16ms binning from the \bat catalog after data standardization). Therefore, we had more refined structures in the \fermi light curves compared to \swift, which can also be a factor for the differences in the embeddings for the two catalogs. The clustering analysis involved approximately 1500 GRBs in the \swift sample and around 3300 GRBs in the \fermi sample, with a small overlap of common GRBs. Given that clustering depends on the overall behavior of each sample, differences between the two samples are expected. Further, the analysis was conducted in the detector count space which might be influenced by the non-linear responses of the instruments. Furthermore, differences in the processing pipelines could lead to classifications being affected by data artifacts and decisions made during processing.

\end{document}